\documentclass[12pt]{amsart}

\usepackage[english]{babel}
\usepackage[T1]{fontenc}
\usepackage[utf8]{inputenc}
\usepackage{epsf,amssymb,amsmath,array,hhline, dsfont}
\usepackage{latexsym}
\usepackage{graphicx}
\usepackage[final]{pdfpages}
\usepackage{verbatim}
\usepackage{stmaryrd}

\title[Random scattering zippers]{Absence of absolutely continuous spectrum for random scattering zippers}

\author[H. Boumaza]{Hakim Boumaza}
\email{boumaza@math.univ-paris13.fr}
\address{Universit\'e Paris 13, Sorbonne Paris Cit\'e\\
LAGA\\
CNRS, UMR 7539\\
F-93430, Villetaneuse\\
France}
\thanks{Research supported by the GDR DYNQUA}

\author[L. Marin]{Laurent Marin}
\email{lrnt.marin@gmail.com}
\address{Institut Fourier,
100 rue des maths, BP 74, 38402 St Martin d'H\`{e}res cedex, France}


\newtheorem{theo}{Theorem}
\newtheorem{defini}{Definition}
\newtheorem{proposi}{Proposition}
\newtheorem{lemma}{Lemma}
\newtheorem{coro}{Corollary}
\newtheorem{rem}{Remark}

\newcommand{\CM}{{\mathbb C}}

\newcommand{\RM}{{\mathbb R}}
\newcommand{\SM}{{\mathbb S}}

\newcommand{\ZM}{{\mathbb Z}}

\newcommand{\CMV}{{\mathbb U}}
\newcommand{\CMVL}{{\mathbb V}}
\newcommand{\CMVR}{{\mathbb W}}

\newcommand{\Ll}{\mathcal{L}}

\newcommand{\one}{\mathds{1}_L}

\newcommand{\inv}{{\mbox{\rm\tiny inv}}}

\newcommand{\UL}{\textrm{U}(L)}

\def\proof{\noindent{\bf Proof:}\ \ }
\def\qed{\hfill $\Box$\medskip}

\begin{document}

%%%%%%%%%%%%%%%%%%%%%%%%%%%%%%%%%%%%%%%%%%%%%%%%%%%%
\begin{abstract}
A scattering zipper is a system obtained by concatenation of scattering events with equal even number of incoming and out going channels. The associated scattering zipper operator is the unitary equivalent of Jacobi matrices with matrix entries. For infinite identical events and random phases, Lyapunov exponents positivity is proved and yields to the absence of absolutely continuous spectrum. 
\end{abstract}
%%%%%%%%%%%%%%%%%%
\maketitle
\vspace{.5cm}

%%%%%%%%%%%%%%%%%%%%%%%%%%%%%%%%%%%%%%%%%%%%%%%%
\section{Random scattering zippers}\label{sec_intro}

\subsection{Scattering zippers}

An infinite scattering zipper describes consecutive scattering events with a fixed number $2L$ of incoming and out-going channels each. It is specified by a sequence $(S_n)_{n \in \ZM}$ of unitary scattering matrices $S_n$ in the unitary group $\mbox{U}(2L)$. Then the scattering zipper operator acting on $\ell^2( \ZM,\CM^L)$ is defined as 
\begin{equation}\label{eq_def_cmv}
\CMV\;=\;\CMVL\,\CMVR\;, 
\end{equation}

\noindent where the two unitaries $\CMVL$ and $\CMVR$ are given by 
$$
\CMVL
\;=\; 
\begin{pmatrix}
\ddots & & & \\
 & S_0 & &  \\
& & S_{2} &  \\
& & & \ddots 
\end{pmatrix}                                                                                            
\;,
\qquad
\CMVR
\;=\;
\begin{pmatrix}
\ddots & & &  \\
& S_{-1}  & &  \\
 & & S_{1}  &  \\
& & & \ddots   
\end{pmatrix}                                                                                            
\;.
$$

\noindent Note that the $2L \times 2L$ blocks in $\CMVL$ are
shifted by $L$ with respect to those of $\CMVR$ along the diagonal.

This model is the matrix-valued generalization of the so-called CMV matrices. Those operators are the unitary analog of Jacobi matrices with matrix coefficients, see \cite{Sim2} and \cite{Sim3} for a large review on that topic. CMV matrices were introduced in the first place to study orthonormal polynomials on the the circle. Later in \cite{CMV}, the nice factorization in two diagonal by block operators we use as a definition was discovered. From the spectral point on view, CMV matrices are in one to one with spectral measure on the circle (Verblunsky Theorem).  Matricial version of CMV matrices were also considered in \cite{DPS}, it corresponds to this model with all the phases set to $\one$, where $\one$ is the identity matrix of order $L$. They do not verify a Verblunsky theorem analog as in scalar case, but scattering zippers do (see \cite{MS} for details). 
\vskip 3mm

Let us now precise the assumptions we will need to make on the operator (\ref{eq_def_cmv}). The sequence $\{S_n\}_{n\in \ZM}$ is chosen to belong to the following subset:
\begin{equation}
\label{eq-U(2L)subset}
\mbox{U}(2L)_\inv
\; =\;
\left\{\left. S = \begin{pmatrix} \alpha & \beta \\ \gamma & \delta \end{pmatrix} \in \mbox{U}(2L), \;\right|\; 
\alpha,\gamma,\delta \in \mathcal{M}_{L}(\CM) \mbox{ and } \beta \in \mathrm{GL}_{L}(\CM)\; 
\right\}
\;.
\end{equation}

Invertibility of upper right entries of size $L\times L$ of the $S_n$'s assures that the scattering is effective so that $\CMV$ does not decouple into a direct sum of two or more parts.

In {\rm \eqref{eq-U(2L)subset}}, equivalent to the condition that $\beta$ is invertible is the condition $\alpha^*\alpha<\one$. Furthermore, one has the representation (see \cite{MS})
\begin{equation}
\label{eq-CMVrep}
\mbox{\rm U}(2L)_\inv
\; =\;
\left\{\left. S(\alpha,U,V) \in \mbox{\rm U}(2L) \;\right|\; \alpha^*\alpha<\one \;\mbox{\rm and }U,V\in\mbox{\rm U}(L) 
\right\}
\;,
\end{equation}
where
\begin{equation}
\label{eq-Smatdef}
S(\alpha,U,V)\; =\; 
\begin{pmatrix} \alpha & \rho(\alpha) U 
\\ 
V\widetilde{\rho}(\alpha) & -V\alpha^*U 
\end{pmatrix}
\;,
\end{equation}
and 
$$
 \rho(\alpha) \;=\; (\one-\alpha\alpha^*)^{\frac{1}{2}}, \;\;\;\; \widetilde{\rho}(\alpha) \;=\; (\one-\alpha^*\alpha)^{\frac{1}{2}}\;.
$$

In what follows, we will simply write $\rho$ and $\widetilde{\rho}$ where there is no ambiguity. The sequence of $\{\alpha_n\}_{n \in \ZM}$ is oftenly called the  Verblunsky sequence associated to $\CMV$.
\vskip 2mm 

\noindent \textbf{Assumption.} We will assume that the Verblunsky sequence is constant, equal to some $\alpha \neq { \bf 0}$. 
\vskip 6mm

\subsection{Random scattering zippers}

We will now introduce the random setting which will allow us to define a random version of the scattering zipper $\CMV$. Let $\Omega_0 = \UL \times \UL$, let $\mathcal{B}_0$ be the Borelian $\sigma$-algebra over $ \UL \times \UL$ for the usual Lie group topology and let $\mathbb{P}_0 \;=\; \nu_L \varotimes \nu_L$ where $\nu_L$ is the Haar measure on $\UL$. Then $\mathbb{P}_0$ is a probability measure on $\Omega_0$. 

We now define the product probability space $(\Omega,\mathcal{B},\mathbb{P})$ :
$$(\Omega,\mathcal{B},\mathbb{P})=\left( (\Omega_0)^{\ZM},\; \varotimes_{n\in \ZM} \mathcal{B}_0,\; \varotimes_{n\in \ZM} \mathbb{P}_0\right).$$
\vskip 2mm

\noindent Let $\omega \in \Omega$. For every $n\in \ZM$, we consider the random unitary matrix $S_n(\omega) \in \mbox{U}(2L)_\inv$ defined by 
$$S_n(\omega) = S(\alpha,U_n(\omega),V_n(\omega)),$$
where $((U_n(\omega),V_n(\omega)))_{n\in \ZM}$ is a sequence of independent and identically distributed (\textit{i.i.d.} for short) random variables on the probability space $(\Omega_0,\mathcal{B}_0,\mathbb{P}_0)$. Then $(S_n(\omega))_{n\in \ZM}$ is a sequence of \textit{i.i.d.} random matrices in $\mbox{U}(2L)_\inv$.

 Associated to this sequence $(S_n(\omega))_{n\in \ZM}$, one can defined as in (\ref{eq_def_cmv}) the operators $\mathbb{V}_{\omega}$, $\mathbb{W}_{\omega}$ and $\mathbb{U}_{\omega}=\mathbb{V}_{\omega}\; \mathbb{W}_{\omega}$. We call \textbf{random scattering zipper} the family of random operators $\{ \mathbb{U}_{\omega} \}_{\omega \in \Omega}$.
\vskip 5mm

\noindent \emph{Notation.} Denote the shift $\tau : \Omega \to \Omega, (\tau(\omega))(2n) \; =\; \omega(2n+2) $. 
\vskip 2mm

\noindent The shift $\tau$ is ergodic on $(\Omega,\mathcal{B},\mathbb{P})$. Moreover, using $\tau$, one show that the family of random operators $\{ \mathbb{U}_{\omega} \}_{\omega \in \Omega}$ is $2\ZM$-ergodic. Thus, there exists $\Sigma \subset \CM$ such that, for $\mathbb{P}$-almost every $\omega \in \Omega$, $\Sigma=\sigma(\mathbb{U}_{\omega})$. There also exist $\Sigma_{\mathrm{pp}}$, $\Sigma_{\mathrm{ac}}$ and $\Sigma_{\mathrm{sc}}$, subsets of $\CM$, such that, for $\mathbb{P}$-almost every $\omega \in \Omega$, $\Sigma_{\mathrm{pp}}=\sigma_{\mathrm{pp}}(\mathbb{U}_{\omega})$, $\Sigma_{\mathrm{ac}}=\sigma_{\mathrm{ac}}(\mathbb{U}_{\omega})$ and $\Sigma_{\mathrm{sc}}=\sigma_{\mathrm{sc}}(\mathbb{U}_{\omega})$, respectively the pure point, absolutely continuous and singular continuous spectrum of $\mathbb{U}_{\omega}$. 
\vskip 2mm

\begin{comment}
\noindent We denote by $\SM^1$ the unit circle in $\CM$ : $\SM^1=\{ z\in \CM \; :\; |z|=1  \}$. Since for every $\omega\in \Omega$, $\sigma(\mathbb{U}_{\omega})=\SM^1$, we have $\Sigma = \SM^1$. \textbf{A VERIFIER}.
\end{comment}
\vskip 6mm

\subsection{Main results}

 Now that we have precisely defined the model we will study in this paper, let us state the two main results we will prove in the next sections. For the definition of the Lyapunov exponents associated to the random scattering zipper $\{ \mathbb{U}_{\omega} \}_{\omega \in \Omega}$, see Section \ref{sec_tm_lyap}.

\begin{theo}\label{thm_pos_Lyap_intro}
 For every $z\in \mathbb{S}^1$, $ \gamma_1(z)>  \gamma_2(z)> \cdots >  \gamma_L(z)> 0.$
\end{theo}
Using Theorem \ref{thm_pos_Lyap_intro}, and adapting some results of Kotani's theory, one deduce the absence of absolutely continuous spectrum for $\{ \mathbb{U}_{\omega} \}_{\omega \in \Omega}$.

\begin{theo}\label{thm_ishii_pastur_intro}
The random scattering zipper $\{ \mathbb{U}_{\omega} \}_{\omega \in \Omega}$ has no absolutely continuous spectrum : $\Sigma_{\mathrm{ac}}=\emptyset$.
\end{theo}
We will actually show a more general result on the multiplicity of the absolutely continuous spectrum of $\mathbb{U}_{\omega}$ related to the number of vanishing Lyapunov exponents (see Theorem \ref{thm_ishii_pastur}), from which we deduce Theorem \ref{thm_ishii_pastur_intro}.
\vskip 5mm

Let us finish this introduction by giving the outline of the article. In Section \ref{sec_tm_lyap} we will present the formalism of transfer matrices for the random scattering zipper and we will define the Lyapunov exponents associated to these transfer matrices. We will also recall some of the first properties of these exponents. In Section \ref{sec_pos_lyap} we will prove Theorem \ref{thm_pos_Lyap_intro} by reducing it to an algebraic result on the Lie group generated by the transfer matrices of  $\{ \mathbb{U}_{\omega} \}_{\omega \in \Omega}$ and then by proving this algebraic result. Finally, in Section \ref{sec_kotani}, we will prove Theorem \ref{thm_ishii_pastur_intro} by adapting many ideas of Kotani and Simon which were found in \cite{KS} to the setting of random scattering zippers. In particular we will characterize the multiplicity of the absolutely continuous spectrum in term of the number of vanishing Lyapunov exponents on a subset of full measure of the unit circle.
\vskip 3mm

Results on the positivity of the Lyapunov exponents are already known for some models of unitary band matrices (see \cite{BHJ,HS}). These unitary models deal with scalar coefficients compared to our model which deals with matrix coefficients. The positivity of the Lyapunov exponents for matrix-valued models is also known for matrix-valued Anderson models, both in the discrete and continuous settings (see \cite{G} for the discrete case and \cite{B} for the continuous case). For all these models, the positivity of the Lyapunov exponents implies absence of absolutely continuous spectrum through the results of Kotani's theory (see \cite{BHJ,KS}).

\vskip 6mm

\section{Transfer matrices and Lyapunov exponents}\label{sec_tm_lyap}

\subsection{Transfer matrices} In this section we will present the formalism of the transfer matrices which allows us to reduce the understanding of the asymptotic behaviour of the  solutions of the equation $\mathbb{U}_{\omega}\phi=z\phi$ (for $z\in \SM^1$) to the understanding of the asymptotic behaviour of a product of random matrices in the Lorentz group.
\vskip 3mm

\noindent Recall that the Lorentz group $\mbox{U}(L,L)$ of signature $(L,L)$ is defined to be the set of $2L\times 2L$ matrices preserving the form
\begin{equation}
\label{eq-Ldef}
\Ll
\;=\;
\begin{pmatrix}
 \one & 0
\\
0 & -\one
\end{pmatrix}
\;.
\end{equation}
Let $z\in \SM^1$. One define the transfer matrices in the following way (see \cite{MS}) : consider the application $T(z,\cdot ): \Omega \to \mathrm{U}(L,L)$, 
$$\forall \omega \in \Omega,\ T(z,\omega) =  \begin{pmatrix} V_{0}(\omega)  & 0 \\ 0 & (U_{0}(\omega))^* \end{pmatrix}
\hat{T}_0(z)
\begin{pmatrix} V_{1}(\omega)  & 0 \\ 0 & (U_{1}(\omega))^* \end{pmatrix}
\hat{T}_1
$$
with
$$
\hat{T}_0(z) = \begin{pmatrix}
z^{-1} \widetilde{\rho}^{-1} & \widetilde{\rho}^{-1}\alpha^* \\
\alpha \widetilde{\rho}^{-1} & z \rho^{-1}
\end{pmatrix}, \qquad
\hat{T}_1 = \begin{pmatrix}
 \widetilde{\rho}^{-1} & \widetilde{\rho}^{-1}\alpha^* \\
\alpha \widetilde{\rho}^{-1} &  \rho^{-1}
\end{pmatrix}.
$$

\begin{defini}\label{def_tm}
Let $z\in \SM^1$, $n\in \ZM$ and $\omega \in \Omega$. The $n$-th transfer matrix associated to the operator $\mathbb{U}_{\omega}$ is the matrix $T(z,\tau^{n}(\omega))$.  
\end{defini}
The sequence $(T(z,\tau^{n}(\omega)))_{n\in \ZM}$ is an \emph{i.i.d.} sequence of random matrices in $\mathrm{U}(L,L)$ because of the \emph{i.i.d.} character of the sequence $((U_n(\omega),V_n(\omega)))_{n\in \ZM}$ in $\mathrm{U}(L)$.
\vskip 2mm

The transfer matrices are obtained directly from the matrices $S_n(\omega)$ through the following bijection between $\mbox{\rm U}(2L)_\inv$ and $\mbox{U}(L,L)$ :
$$
\varphi\ :\ \begin{array}{ccl}
            \mbox{\rm U}(2L)_\inv & \to & \mathrm{U}(L,L) \\[2mm]
	    \left( \begin{matrix}
	    \alpha & \beta \\
	    \gamma & \delta
	    \end{matrix} \right) & \mapsto & \left(\begin{matrix}
	    \gamma-\delta \beta^{-1} \alpha & \delta \beta^{-1} \\
	    -\beta^{-1} \alpha & \beta^{-1}
	    \end{matrix} \right).
            \end{array}
$$
Then we have the relation (see \cite{MS}): 
$$\forall \omega \in \Omega,\ \forall z\in \SM^1,\ \forall k\in \ZM,\ T(z,\tau^{2k}(\omega))= \varphi(z^{-1}S_{2k}(\omega))\cdot \varphi(S_{2k+1}(\omega)).$$

\vskip 3mm
To define the Lyapunov exponents in the next section, we need to remark that the transfer matrices $T(z,\cdot)$ generate the cocycle $\Phi$ on the ergodic dynamical system $(\Omega, \mathcal{B}, \mathbb{P}, (\tau^n)_{n \in \ZM} )$ with  $\Phi(z,\cdot,\cdot): \Omega \times \ZM \to \mathrm{U}(L,L) $
$$
\Phi(z,\omega,n) \;=\; \left\lbrace 
\begin{array}{lcl}
T(z,\tau^{n-1}(\omega)) \dots T(z,\omega) & \mbox{ if } & n >0 \\
\mathds{1}_{2L} & \mbox{ if } & n=0 \\
(T(z,\tau^{n}(\omega))^{-1} \dots (T(z,\tau^{-1}(\omega))^{-1} & \mbox{ if } & n <0.
\end{array} \right.
$$

\noindent \textbf{Remark.} For fixed $z\in \SM^1$ and $\omega \in \Omega$, the asymptotic behaviour of the  solutions of $\mathbb{U}_{\omega}\phi=z\phi$ is related to the asymptotic behaviour of the sequence $(||\Phi(z,\omega,n)||)_{n\in \ZM}$ where $||\; ||$ is any norm on $\mathrm{U}(L,L)$.

\vskip 6mm

\subsection{Lyapunov exponents} Using the cocycle $\Phi$, we can now define the Lyapunov exponents associated to the family $\{ \mathbb{U}_{\omega} \}_{\omega \in \Omega}$.

\begin{proposi}\label{prop_lim_def_lyap}
 Let $z\in \SM^1$. For $\mathbb{P}$-almost every $\omega \in \Omega$, the following limits exist :
\begin{equation}\label{eq_lim_Lyap}
 \Psi(z,\omega) := \lim_{n\to +\infty} ((\Phi(z,\omega,n))^*\Phi(z,\omega,n))^{1/2n} = \lim_{n\to -\infty} ((\Phi(z,\omega,n))^*\Phi(z,\omega,n))^{1/2|n|}.
\end{equation}
For every $k\in \{ 1,\ldots, 2L\}$, we denote by $\lambda_k(z,\omega)$ the eigenvalues of $\Psi(z,\omega)$ arranged in decreasing order. Then, there exist real numbers $\lambda_k(z) \geq 0$ such that for $\mathbb{P}$-almost every $\omega \in \Omega$, $\lambda_k(z,\omega)=\lambda_k(z)$.
\end{proposi}

\proof This is a direct consequence of Oseledets theorem applied to the cocycle $\Phi(z,\cdot,\cdot)$. Indeed, according to \cite[Remark 3.4.10]{Arn}, one can apply \cite[Theorem 3.4.11]{Arn} on $\CM^{2L}$ instead of $\RM^{2L}$. 
\qed

\noindent \textbf{Notation.} We denote by $\Omega_{\mbox{Lyap}}$ a subset of $\Omega$ such that $\mathbb{P}(\Omega_{\mbox{Lyap}})=1$ and, for every $\omega\in \Omega_{\mbox{Lyap}}$ the limits (\ref{eq_lim_Lyap}) exist and, for every $k\in \{1,\ldots,2L\}$, $\lambda_k(z,\omega)=\lambda_k(z)$.\\

\noindent Proposition \ref{prop_lim_def_lyap} leads to the definition of the Lyapunov exponents associated to the operator $\mathbb{U}_{\omega}$.

\begin{defini}\label{prop_def_lyap}
We define the $2L$ Lyapunov exponents associated to the family $\{ \mathbb{U}_{\omega} \}_{\omega \in \Omega}$ by : 
$$ \forall \ k\in \{ 1,\ldots,2L\},\ \gamma_k(z) := \log( \lambda_k(z) ).$$  
\end{defini}

Using the fact that the transfer matrices lie in the group $\mathrm{U}(L,L)$ and thus for every $n\in \ZM$ the matrix $\Phi(z,\omega,n)$ lies in this group too, we get a symmetry relation for the Lyapunov exponents :
$$\forall j\ \in \{0,\ldots L \},\ \gamma_{2L-j+1}(z)=-\gamma_{j}(z).$$
It implies that the Lyapunov exponents arrange by pairs of opposite real numbers : 
\begin{equation}\label{eq_lyap_pairs}
  \gamma_1(z)\geq  \gamma_2(z)\geq \cdots \geq  \gamma_L(z)\geq 0 \geq -\gamma_L(z) \geq \cdots \geq -\gamma_1(z).
\end{equation}
Thus we will only have to consider the $L$ first Lyapunov exponents which are positive numbers : 
$$ \gamma_1(z)\geq  \gamma_2(z)\geq \cdots \geq  \gamma_L(z)\geq 0.$$

\noindent In the next section, we study the Lyapunov exponents associated to the cocycle $\Phi$ and we prove that they are all distincts and thus, using (\ref{eq_lyap_pairs}) $\gamma_1(z),\ldots,\gamma_L(z)$ are all positive. 

\vskip 6mm

\section{Positivity of the Lyapunov exponents}\label{sec_pos_lyap}

\subsection{Reduction of Theorem \ref{thm_pos_Lyap_intro} to an algebraic result}\label{sec_pos_lyap_thm}

This section is devoted to the proof of the positivity of the Lyapunov exponents of the family $\{ \mathbb{U}_{\omega} \}_{\omega \in \Omega}$. We will explain how to reduce the proof of Theorem \ref{thm_pos_Lyap_intro} to an algebraic result on some subgroup of the Lie group $\mathrm{U}(L,L)$.
\vskip 3mm

To prove that the Lyapunov exponents are all distinct and strictly positive, we need to consider the F\"{u}rstenberg group associated to $\mathbb{U}_{\omega}$, which is the subgroup of $\mathrm{U}(L,L)$ generated by all the transfer matrices :
$$
G(z) \; = \; \overline {\langle  \text{supp} \mu_z  \rangle} \subset \mathrm{U}(L,L)
$$
where $\mu_z$ is the common law of all the transfer matrices $T(z,\tau^n(\omega))$, for every $n\in \ZM$. The closure is taken for the usual topology on $\mathrm{U}(L,L)$ which is the topology induced on $\mathrm{U}(L,L)$ by the usual topology on $\mathcal{M}_{2L}(\CM)$.
\vskip 2mm

Directly from definition and using the fact that $(T(z,\tau^{n}(\omega)))_{n\in \ZM}$ is an \emph{i.i.d.} sequence of random matrices in $\mathrm{U}(L,L)$, we get the following internal description of $G(z)$ : 
$$
G(z) \; = \; \overline{ \left\{ \left\langle  \left(\begin{smallmatrix} V_{0} & 0 \\ 0 & (U_{0})^* \end{smallmatrix} \right)
\hat{T}_0(z)
 \left(\begin{smallmatrix} V_{1}  & 0 \\ 0 & (U_{1})^* \end{smallmatrix}\right)
\hat{T}_1  \Big| (U_0,V_0,U_1,V_1) \in \mathrm{U}(L)^4  \right\rangle \right\} } \subset \mathrm{U}(L,L).
$$
We will use this description of the elements of $G(z)$ to prove in Proposition \ref{prop_G_U}  that actually the subgroup $G(z)$ is the whole group $\mathrm{U}(L,L)$. Taking in account the result of Proposition \ref{prop_G_U}, we can prove Theorem \ref{thm_pos_Lyap_intro}.

\proof Assuming that for every $z\in \SM^1$, $G(z)=\mathrm{U}(L,L)$, we can follow the strategy of \cite[Theorem 6.1]{ABJ}. We fix $z\in \SM^1$. By Cayley transform, the group $G(z)$ is unitarily equivalent to the complex symplectic group. Indeed, if we denote by $C\in \mathcal{M}_{2L}(\CM)$ the matrix
$$C=\frac{1}{\sqrt{2}} \left( \begin{matrix}
                              \one & -i\one \\
			      \one & i\one
                              \end{matrix}  \right),$$
and by $J$ the matrix 
$$J= \left( \begin{matrix}
            0 & -\one \\
	    \one & 0
            \end{matrix}  \right),$$
we have
$$\mathrm{U}(L,L) = C \mathrm{Sp}_{\mathrm{L}}(\CM) C^*$$
where
$$\mathrm{Sp}_{\mathrm{L}}(\CM) = \{ M\in \mathcal{M}_{2L}(\CM) | M^* J M = J \}.$$
\vskip 2mm
Since the results of \cite{BL} are stated for the real symplectic group, we introduce, as in \cite{ABJ}, the following application which split the real and imaginery parts of the matrices in $\mathcal{M}_{2L}(\CM)$ 
$$\tau\ :\ \begin{array}{cll}
           \mathcal{M}_{2L}(\CM) & \to & \mathcal{M}_{4L}(\RM) \\[2mm]
	    A+iB & \mapsto & \left( \begin{array}{cc}
				      A & -B \\
				      B & A
				      \end{array} \right) .
				\end{array}
$$
So we have $\tau(C^* \mathrm{U}(L,L) C) \subset \mathrm{Sp}_{\mathrm{2L}}(\RM)$. With this setting, we deduce immediately from  \cite[Lemma 6.3]{ABJ} that  $\tau(C^* \mathrm{U}(L,L) C)$ is $L_{2p}$-strongly irreducible for every $p\in \{1,\ldots,L\}$ (for a definition of the notion $L_p$-strong irreducibility, see \cite[Definition A.IV.3.3]{BL}). Then, to adapt the proof of \cite[Lemma 6.4]{ABJ} to our setting, we only need to perform a permutation of lines and columns of the matrix with $2L$ distinct singular values they define. More precisely, let $P$ denote the permutation matrix which send, for $j\in \{1,\ldots, L\}$, the $(L+j)$-th line of the identity matrix of order $2L$ to the line $2j$ and, for $k\in \{0,\ldots, L-1\}$, the $k$-th line to the line $2k+1$. Then, multiplying on left and on right by $P$ the matrix they used in their proof, we obtain an element of $\mathrm{U}(L,L)$ with $2L$ distinct singular values. Thus  $\tau(C^* \cdot \mathrm{U}(L,L)\cdot C)$ is $2p$-contracting for every $p\in \{1,\ldots,L\}$ (for a definition of $p$-contractivity, see \cite[Definition A.IV.1.1]{BL}). 
\vskip 2mm

Now we can apply \cite[Proposition A.IV.3.4]{BL} to the group $\tau(C^*  \mathrm{U}(L,L) C)$ to obtain that all Lyapunov exponents are distincts. Since we have the inequalities (\ref{eq_lyap_pairs}), the $L$ first Lyapunov exponents are strictly positive, which finishes the proof.
\qed

\vskip 6mm

\subsection{Proof of Proposition \ref{prop_G_U}}\label{sec_GU}

In this section, we prove the algebraic result on which the proof of Theorem \ref{thm_pos_Lyap_intro} is based on.

\begin{proposi}\label{prop_G_U}
For any $z \in  \SM^1$, $G(z)  \;=\; \mathrm{U}(L,L)$.
\end{proposi}

\proof
By connexity of $\mathrm{U}(L,L)$, it is enough to show that the Lie algebras of $G(z)$ and $\mathrm{U}(L,L)$ are equal. 
The Lie algebra associated to $\mathrm{U}(L,L)$ is given by
\begin{align*}
\text{Lie} (\mathrm{U}(L,L)) &= \mathfrak{u}(L,L) = \{ T \in \mathcal{M}_{2L} (\CM) \;|\; T^*\Ll + \Ll T =0 \} \\
 &= \left\{ \begin{pmatrix} A & B \\ B^* & D \end{pmatrix} \in \mathcal{M}_{2L} (\CM) \;\Big|\; A^*= -A, D^* = -D, (A,B,D) \in  \mathcal{M}_{L} (\CM) \right\}
\end{align*}

We denote Lie($G(z)$)  =  $\mathfrak{g}(z)$. Now, the proof divides in numerous small Lemma. By taking in account Lemma \ref{base1} and Lemma \ref{base2}, one gets that $\mathfrak{u}(L,L)= \mathfrak{a}_1 \oplus \mathfrak{a}_2 \subset \mathfrak{g}(z)$ and thus $\mathfrak{u}(L,L)=\mathfrak{g}(z)$ which ends the proof.
\qed

\begin{lemma}
For any $(U_0,V_0)\in \mathrm{U}(L)^2$, the matrix $\begin{pmatrix} V_0 & 0 \\ 0 & U_0^* \end{pmatrix} $ belongs to $G(z)$.
\end{lemma}

\proof
Setting $U_0 = V_0 = U_1 = V_1 = \mathds{1}_L$, one gets $\hat{T}_0(z)\hat{T}_1 \in G(z)$. As $G(z)$ is a multiplicative group,   $(\hat{T}_0(z)\hat{T}_1)^{-1}$ also belongs to $G(z)$. Thus only setting $ U_1 = V_1 = \mathds{1}_L$ and $U_0, V_0$ free and multiplying by $(\hat{T}_0(z)\hat{T}_1)^{-1}$ yields to the statement.
\qed

\begin{lemma}\label{base1}
The algebra $\mathfrak{g}(z)$ contains $\mathfrak{a}_1 =\left\{ \begin{pmatrix} A & 0 \\ 0 & D \end{pmatrix} \Big| A^* = -A, D^*= -D, (A,D) \in \mathcal{M}_L (\CM)^2 \right\}$.
\end{lemma}

\proof
By previous Lemma, one has $\mathrm{U}(L) \varoplus \mathrm{U}(L) \subset G(z)$. Therefore this implies $\text{Lie} (\mathrm{U}(L)) \varoplus \text{Lie} (\mathrm{U}(L)) \subset \mathfrak{g}(z)$.
As $\text{Lie} (\mathrm{U}(L)) = \{ A \in \mathcal{M}_L (\CM) | A^* = -A \}$, this implies the statement.
\qed

\begin{lemma}
The matrix $i \begin{pmatrix} 0 & \widetilde{\rho}^{-2} \alpha^* \\  \alpha \widetilde{\rho}^{-2} & 0 \end{pmatrix}$ belongs  to $\mathfrak{g}(z)$.
\end{lemma}

\proof
For $V_0 = U_0 = \mathds{1}_L$,  $ \hat{T}_0(z) \begin{pmatrix} V_1 & 0 \\ 0 & U_1^* \end{pmatrix} \hat{T}_1 \in G(z)$.

Therefore, $\hat{T}_1^{-1}  \begin{pmatrix} V_1 & 0 \\ 0 & U_1^* \end{pmatrix} \hat{T}_1 \; =  \; (\hat{T}_0(z)\hat{T}_1)^{-1}\hat{T}_0(z) \begin{pmatrix} V_1 & 0 \\ 0 & U_1^* \end{pmatrix} \hat{T}_1 \in G(z)$.

Now for $j \in [\![1,L]\!]$ and any $t \in \RM$, picking $V_1$ to be the matrix $\mbox{diag}(1,\ldots,1,e^{it},1,\ldots,1)$ with $e^{it}$ at the $j$-th place, and $U_1=\mathds{1}_L$
%$\begin{pmatrix} 
%1 &  & & & & &  \\ 
%& \ddots & & & 0 & \\
%& & 1 & & & & \\
%& & & e^{it} & & & \\
%& & & &  1& & \\
%& 0 & & & & \ddots & \\
%& & & & & & 1
%\end{pmatrix}$ 
and derivating in $t=0$ yields to 
$$
\forall j \in [\![1,L]\!], i \hat{T}_1^{-1}  \begin{pmatrix} E_{jj} & 0 \\ 0 & 0 \end{pmatrix} \hat{T}_1 \in \mathfrak{g}(z).
$$
In particular summing over $j$ gives 
$$
 i \hat{T}_1^{-1}  \begin{pmatrix} \mathds{1}_L & 0 \\ 0 & 0 \end{pmatrix} \hat{T}_1 \in \mathfrak{g}(z)
$$
and the result writing $\hat{T}_1$  as block matrix.
\qed

\begin{lemma}\label{lemma4}
There exists an index $(j_0,k_0) \in [\![1,L]\!]^2$ and $c \in \CM,$ with $c\neq 0 $ such that 
$$
i \begin{pmatrix} 0 & c E_{k_0j_0} \\ \overline{c} E_{j_0k_0} & 0 \end{pmatrix}  \in \mathfrak{g}(z).
$$  
\end{lemma}

\proof
As we suppose $\alpha \neq 0$ and $\widetilde{\rho}^{-2}$ is invertible then $\widetilde{\rho}^{-2}\alpha^* \neq 0$. So there exists an index $(j_0,k_0) \in [\![1,L]\!]^2$ such that $(\widetilde{\rho}^{-2}\alpha^*)_{(j_0,k_0)} = c  \neq 0$ and 
$$
\left[ i\begin{pmatrix} 0 & 0\\0 & E_{j_0j_0} \end{pmatrix},
\left[ i\begin{pmatrix} E_{k_0k_0} & 0 \\ 0 & 0\end{pmatrix},
i\begin{pmatrix} 0 & \widetilde{\rho}^{-2}\alpha^*\\ \alpha \widetilde{\rho}^{-2} & 0\end{pmatrix}\right] \right] \;=\;
 i\begin{pmatrix} 0 & cE_{k_0j_0}\\ \overline{c} E_{j_0k_0} & 0\end{pmatrix}
\in \mathfrak{g}(z).
$$
\qed

\begin{lemma}\label{base2}
$\mathfrak{a}_2 = \left\{ \begin{pmatrix} 0 & B \\ B^* & 0 \end{pmatrix} \Big| B \in \mathcal{M}_L (\CM) \right\}  \subset \mathfrak{g}(z).$
\end{lemma}

\proof
It is enough to show that the elements $\left( \begin{smallmatrix} 0 & y E_{jk} \\ \overline{y} E_{kj} & 0 \end{smallmatrix} \right)$ for $(j,k) \in [\![1,L]\!]^2$, $y=1$ and $y=i$ belong to $\mathfrak{g}(z)$, as they form a basis of $\mathfrak{a}_2$.
From Lemma \ref{base1}, the matrix $ \left( \begin{smallmatrix} yE_{jk} - \overline{y} E_{kj} & 0 \\ 0 & 0 \end{smallmatrix} \right)$ belongs to $\mathfrak{g}(z)$ as the block $ yE_{jk} - \overline{y} E_{kj}$ is anti-hermitian for any $(j,k) \in [\![1,L]\!]^2$ and any $y \in \CM$.
For any $k \in [\![1,L]\!]^2, k \neq k_0$ and any $y \in \CM$ one has, 
$$
\left[ \begin{pmatrix}yE_{kk_0}-\overline{y} E_{k_0k}  & 0\\0 & 0\end{pmatrix},
 i\begin{pmatrix} 0 & cE_{k_0j_0}\\ \overline{c} E_{j_0k_0} & 0\end{pmatrix}\right]  = i\begin{pmatrix} 0 & ycE_{kj_0} \\ -\overline{yc} E_{j_0k} & 0  \end{pmatrix} \in \mathfrak{g}(z).
$$
%and 
%$$
%\left[ \begin{pmatrix} 0  & 0\\0 & yE_{j_0,j}-\overline{y} E_{j,j_0}\end{pmatrix},
% i\begin{pmatrix} 0 & cE_{k_0,j_0}\\ \overline{c} E_{j_0,k_0} & 0\end{pmatrix}\right]  = i\begin{pmatrix} 0 & ycE_{k_0,j} \\ \overline{yc} E_{j,k_0} & 0  \end{pmatrix} \in \mathfrak{g}^z
%$$
%Indeed, as $y$ varies in the complex plane, one obtains basis elements on the $j_0$-row and $k_0$-line. Doubling the same computation give then any elements of the basis, except the one in $(k_0,j_0)$. 
Then, for any $j \in [\![1,L]\!]^2, j \neq j_0$,
$$
\left[  i\begin{pmatrix} 0 & ycE_{kj_0} \\ -\overline{yc} E_{j_0k} & 0  \end{pmatrix}, i\begin{pmatrix} 0 & 0\\ & E_{jj_0}+E_{j_0j}\end{pmatrix}\right]  = \begin{pmatrix} 0 & -ycE_{kj} \\ -\overline{yc} E_{jk} & 0  \end{pmatrix} \in \mathfrak{g}(z).
$$
One can choose either $y=c^{-1}$ or $y=ic^{-1}$ to get all the elements of the basis of $\mathfrak{a}_2$ except the two corresponding to the case $j = j_0, k= k_0$. In this case, one obtains from the same computation that
$$
i\begin{pmatrix} 0 & (y-\overline{y})cE_{k_0j_0} \\ (\overline{y}-y)\overline{c} E_{j_0k_0} & 0  \end{pmatrix} \in \mathfrak{g}(z).
$$
Since $(y-\overline{y})$ is purely imaginary, real linear combinations with the element given in Lemma \ref{lemma4} complete the basis of the set $\mathfrak{a}_2$.
\qed

\begin{rem}
In case $\alpha ={\bf 0} $, one can show that $ \mathfrak{g}(z) = \mathfrak{a}_1 \subsetneq \mathfrak{u}(L,L)$. This justifies our hypothesis $\alpha \neq {\bf 0} $.
\end{rem}

\vskip 6mm

\section{Absence of absolutely continuous spectrum}\label{sec_kotani}

In this section, we will prove that Theorem \ref{thm_pos_Lyap_intro} implies absence of absolutely continuous spectrum for the family $\{\mathbb{U}_{\omega} \}_{\omega\in \Omega}$. For this purpose, we will prove an analog of the theorem of Ishii and Pastur on the characterization of the absolutely continuous spectrum in term of zeros of the Lyapunov exponents. More precisely, we will obtain a unitary version of the matrix-valued analog of Ishii-Pastur's theorem which was proven in \cite[Theorem 5.4]{KS}.

We recall that for $\omega \in \Omega_{\mbox{Lyap}}$, the operator $\mathbb{U}_{\omega}$ has $2L$ Lyapunov exponents which can be regrouped by pairs of opposite real numbers : 
$$\gamma_1(z) \geq \cdots \geq \gamma_L(z)\geq 0 \geq \gamma_{L+1}(z)=-\gamma_L(z) \geq \cdots \geq \gamma_{2L}(z)=-\gamma_1(z).$$

For $j\in \{1,\ldots,L\}$, we set 
$$Z_j=\{z\in \SM^1\ |\ \mbox{there exist } l_1,\ldots, l_{2j}\in \{1,\ldots,2L \},\ \gamma_{l_1}(z)=\cdots=\gamma_{l_{2j}}(z)=0 \}.$$
We also recall that a sequence $\varphi\in (\CM^{L})^{\ZM}$ is said to be \emph{polynomially bounded} if there exist $C>0$ and $p\geq 1$ such that
$$\forall n\in \ZM,\ ||\varphi_n||_{\CM^L} \leq C(1+|n|)^p,$$
where $||\cdot ||_{\CM^L}$ is any norm on $\CM^L$. Now we can state the first result of this section.

\begin{proposi}\label{prop_ishii_1}
 Let $j\in \{1,\ldots,L\}$ and let $z\in Z_j$ be fixed. Let $\omega\in \Omega_{\mbox{Lyap}}$. Then, every subspace of 
$\{ \varphi\in (\CM^{L})^{\ZM}\ |\ \mathbb{U}_{\omega}\varphi = z \varphi,\ \varphi \notin \ell^2(\ZM)\otimes \CM^{L}\mbox{ and } \varphi \mbox{ is polynomially bounded } \}$ has dimension at most $2j$.
\end{proposi}

\proof
We set 
$$V_{\mathrm{sol}}(z)=\{ \varphi\in (\CM^{L})^{\ZM}\ |\ \mathbb{U}_{\omega}\varphi = z \varphi \},$$

$$V_{\mathrm{P}}(z)=\{ \varphi\in V_{\mathrm{sol}}(z)\ |\ \varphi \mbox{ is polynomially bounded } \},$$
and
$$V_{\ell^2}(z)=\{ \varphi\in V_{\mathrm{sol}}(z)\ |\ \varphi  \in \ell^2(\ZM)\otimes \CM^L \}. $$
To prove the proposition, we have to show that
$$\mbox{dim }V_{\mathrm{P}}(z) \leq 2j + \mbox{dim }V_{\ell^2}(z). $$ 
For $\varphi=(\varphi_n)_{n\in \ZM}$ and  $\psi=(\psi_n)_{n\in \ZM}$ in $V_{\mathrm{sol}}(z)$, we set 
$$W(\varphi,\psi)=\left(\begin{array}{c}
                        \varphi_0 \\
			\varphi_1
                        \end{array} \right)^* \mathcal{L}\left(\begin{array}{c}
                        \psi_0 \\
			\psi_1
                        \end{array} \right). $$
Then, $W$ is an antisymmetric form on $V_{\mathrm{sol}}(z)$. Moreover, since $\psi$ is in $V_{\mathrm{sol}}(z)$, it is uniquely determined by $\left(\begin{smallmatrix}
\psi_0 \\
\psi_1                                                                                                                                                  
\end{smallmatrix} \right)$ and since $\mathcal{L}$ is non-singular, $W$ is non-degenerate. Indeed, let $\psi\in V_{\mathrm{sol}}(z)$ be such that, for every $\varphi \in V_{\mathrm{sol}}(z)$, $W(\varphi,\psi)=0$. Then, we apply this equality with $2L$ different $\varphi$'s, those being such that $\left(\begin{smallmatrix}
\varphi_0 \\
\varphi_1                                                                                                                                                  
\end{smallmatrix} \right)$ are the $2L$ vectors of the canonical basis of $\CM^{2L}$. Writing these $2L$ relations, we obtain a Cramer system with unique solution $\mathcal{L}\left(\begin{smallmatrix}
\psi_0 \\
\psi_1                                                                                                                                                  
\end{smallmatrix} \right) =0$. Since $\mathcal{L}$ is non-singular, we get $\left(\begin{smallmatrix}
\psi_0 \\
\psi_1                                                                                                                                                  
\end{smallmatrix} \right)=0$ and  since $\psi$ is uniquely determined by $\left(\begin{smallmatrix}
\psi_0 \\
\psi_1                                                                                                                                                  
\end{smallmatrix} \right)$, we get $\psi=0$.
\vskip 2mm

\noindent Now, since $W$ is a non-degenerate antisymmetric form on $V_{\mathrm{sol}}(z)$, if $V_1$ and $V_2$ are two subspaces of $V_{\mathrm{sol}}(z)$ such that
\begin{equation}\label{eq_W_orth}
\forall \varphi\in V_1,\ \forall \psi\in V_2,\ W(\varphi,\psi)=0, 
\end{equation}
then 
\begin{equation}\label{eq_W_orth_dim}
\mbox{dim }V_1+\mbox{dim }V_2 \leq 2L. 
\end{equation}
Indeed, $V_1$ and $\mathcal{L}V_2$ are orthogonal for $W$. 

\noindent We set
$$D_{\pm} = \{\varphi\in V_{\mathrm{sol}}(z)\ |\ \varphi \mbox{ decays exponentially at } \pm\infty \}.$$
Since $z\in Z_j$, exactly $2j$ among the Lyapunov exponents vanish at $z$. Thus, by Oseledets theorem, 
$$\mbox{dim } D_{\pm} = L-j.$$
We also have $D_+\cap D_- \subset V_{\ell^2}(z)$, so 
\begin{equation}\label{eq_dim_D}
\mbox{dim }(D_+ + D_-)= \mbox{dim }D_+ +  \mbox{dim }D_- - \mbox{dim }(D_+ \cap D_-) \geq 2L-2j-\mbox{dim }V_{\ell^2}(z)
\end{equation}
Moreover, if we take $\varphi \in V_{\mathrm{P}}(z)$ and $\psi\in D_+ +D_-$, then, by direct domination, we have
\begin{equation}\label{eq_lim_D}
 \lim_{|n|\to +\infty} \left(\begin{array}{c}
                        \varphi_n \\
			\varphi_{n+1}
                        \end{array} \right)^* \mathcal{L}\left(\begin{array}{c}
                        \psi_n \\
			\psi_{n+1}
                        \end{array} \right) =0.
\end{equation}
But, one can actually show that the sequence $\left( \left(\begin{smallmatrix}
                        \varphi_n \\
			\varphi_{n+1}
                        \end{smallmatrix}\right)^* \mathcal{L} \left(\begin{smallmatrix}
                        \psi_n \\
			\psi_{n+1}
                        \end{smallmatrix}\right) \right)_{n\in \ZM}$ is constant when we choose $\varphi$ and $\psi$ in  $V_{\mathrm{sol}}(z)$.
Indeed, since $|z|=1$, $T(z,\tau^{n}(\omega))\in \mathrm{U}(L,L)$ for every $n\in \ZM$. Thus, 
\begin{eqnarray}
\forall n\in \ZM,\  \left(\begin{smallmatrix}
                        \varphi_{n+1} \\
			\varphi_{n+2}
                        \end{smallmatrix} \right)^* \mathcal{L}\left(\begin{smallmatrix}
                        \psi_{n+1} \\
			\psi_{n+2}
                        \end{smallmatrix} \right) & = &  \left( T(z,\tau^{n+1}(\omega)) \left(\begin{smallmatrix}
                        \varphi_{n} \\
			\varphi_{n+1}
                        \end{smallmatrix}\right) \right)^* \mathcal{L}T(z,\tau^{n+1}(\omega))\left( \begin{smallmatrix}
                        \psi_n \\
			\psi_{n+1}
                        \end{smallmatrix} \right) \nonumber \\
 & = & \left(\begin{smallmatrix}
                        \varphi_{n} \\
			\varphi_{n+1}
                        \end{smallmatrix} \right)^* T(z,\tau^{n+1}(\omega))^* \mathcal{L}T(z,\tau^{n+1}(\omega))\left( \begin{smallmatrix}
                        \psi_n \\
			\psi_{n+1}
                        \end{smallmatrix} \right) \nonumber \\
& = &  \left(\begin{smallmatrix}
                        \varphi_n \\
			\varphi_{n+1}
                        \end{smallmatrix} \right)^* \mathcal{L}\left(\begin{smallmatrix}
                        \psi_n \\
			\psi_{n+1}
                        \end{smallmatrix} \right). \nonumber
\end{eqnarray}
In particular,
\begin{equation}\label{eq_lim_D_2}
 \left(\begin{smallmatrix}
                        \varphi_0 \\
			\varphi_1
                        \end{smallmatrix} \right)^* \mathcal{L}\left(\begin{smallmatrix}
                        \psi_0 \\
			\psi_1
                        \end{smallmatrix} \right) = \lim_{|n|\to +\infty}  \left(\begin{smallmatrix}
                        \varphi_n \\
			\varphi_{n+1}
                        \end{smallmatrix} \right)^* \mathcal{L}\left(\begin{smallmatrix}
                        \psi_n \\
			\psi_{n+1}
                        \end{smallmatrix} \right) = 0.
\end{equation}
Thus, if  $\varphi \in V_{\mathrm{P}}(z)$ and $\psi\in D_+ +D_-$, $W(\varphi,\psi)=0$. Then, by (\ref{eq_W_orth_dim}),
$$\mbox{dim }V_{\mathrm{P}}(z) +\mbox{dim }(D_+ +D_-)\leq 2L.$$
Combining this inequality with (\ref{eq_dim_D}), one finally get
$$\mbox{dim }V_{\mathrm{P}}(z)\leq 2j +  \mbox{dim }V_{\ell^2}(z), $$
which proves the proposition.
\qed

We introduce the set $\mathrm{P}_{\mathrm{bdd}}$ of all the complex numbers $z$ of modulus $1$ such that the equation $\mathbb{U}_{\omega}\varphi = z\varphi$ admits a non-trivial polynomially bounded solution. Since our operator $\mathbb{U}_{\omega}$ has band structure, we can use results from \cite{BHJ}. In particular, \cite[Lemma 5.4]{BHJ} and \cite[Corollary 5.2]{BHJ} applied to our model lead to the following result.

\begin{proposi}\label{prop_spec_P}
For $\mathbb{P}$-almost every $\omega\in \Omega$,
$$\sigma(\mathbb{U}_{\omega}) = \overline{\mathrm{P}_{\mathrm{bdd}}}=\overline{\{ z\in \SM^1 \ |\ \exists \varphi\in V_{\mathrm{P}}(z)\setminus \{0\} \}}$$
and $\mathrm{E}_{\SM^1 \setminus \mathrm{P}_{\mathrm{bdd}}}(\mathbb{U}_{\omega})=0$, where $\mathrm{E}_{\SM^1 \setminus \mathrm{P}_{\mathrm{bdd}}}(\mathbb{U}_{\omega})$ is the spectral projector on $\SM^1 \setminus \mathrm{P}_{\mathrm{bdd}}$ associated to the unitary operator $\mathbb{U}_{\omega}$.
\end{proposi}

With this proposition, we can finally prove the main result of this section, an Ishii-Pastur theorem for our model.

\begin{theo}\label{thm_ishii_pastur}
 For every $\omega \in \Omega_{\mathrm{Lyap}}$, the multiplicity of the absolutely continuous spectrum of $\mathbb{U}_{\omega}$ on $Z_j$ is at most $2j$.
\end{theo}

\proof
Let $\omega \in \Omega_{\mathrm{Lyap}}$. For $\Delta$ a Borelian subset of $\SM^1$, we denote by $\mathrm{E}_{\Delta}(\mathbb{U}_{\omega})$ the spectral projection on $\Delta$ and by  $\mathrm{E}_{\Delta}^{\mathrm{a.c}}(\mathbb{U}_{\omega})$ the spectral projection on the absolutely continuous part of $\sigma(\mathbb{U}_{\omega})$ in $\Delta$.

\noindent To prove the theorem, we have to prove that
$$\mathrm{rk}\ \mathrm{E}_{Z_j}^{\mathrm{a.c}}(\mathbb{U}_{\omega}) \leq 2j.$$
Since by Proposition \ref{prop_spec_P}, $\mathrm{E}_{\SM^1 \setminus \mathrm{P}_{\mathrm{bdd}}}(\mathbb{U}_{\omega})=0$, we have
$$ \mathrm{E}_{Z_j}^{\mathrm{a.c}}(\mathbb{U}_{\omega}) = \mathrm{E}_{Z_j \cap \mathrm{P}_{\mathrm{bdd}}}^{\mathrm{a.c}}(\mathbb{U}_{\omega}) =  \mathrm{E}_{Z_j \cap \mathrm{P}_{\mathrm{bdd}}\cap S}^{\mathrm{a.c}}(\mathbb{U}_{\omega}) +  \mathrm{E}_{Z_j \cap \mathrm{P}_{\mathrm{bdd}}\cap S^{c}}^{\mathrm{a.c}}(\mathbb{U}_{\omega}),$$
where $S=\{ z\in \SM^1 \ |\ \exists \varphi \in V_{\mathrm{P}}(z) \cap V_{\ell^2}(z),\ \varphi \neq 0 \}$. If $z\in S$, then $z$ is an eigenvalue of $\mathbb{U}_{\omega}$. Since there are only countably many $\ell^2$-eigenvectors for $\mathbb{U}_{\omega}$, the Haar measure of $S$ is zero. Thus, 
\begin{equation}\label{eq_ishii_1}
\mathrm{E}_{Z_j \cap \mathrm{P}_{\mathrm{bdd}}\cap S}^{\mathrm{a.c}}(\mathbb{U}_{\omega})=0\quad \mbox{and}\quad  \mathrm{E}_{Z_j}^{\mathrm{a.c}}(\mathbb{U}_{\omega}) =  \mathrm{E}_{Z_j \cap \mathrm{P}_{\mathrm{bdd}}\cap S^{c}}^{\mathrm{a.c}}(\mathbb{U}_{\omega}).
\end{equation}
But, since $z\in Z_j$, we can apply Proposition \ref{prop_ishii_1} to get directly that
$$\mathrm{rk}\   \mathrm{E}_{Z_j \cap \mathrm{P}_{\mathrm{bdd}}\cap S^{c}}(\mathbb{U}_{\omega}) \leq 2j $$
which implies
\begin{equation}\label{eq_ishii_2}
\mathrm{rk}\    \mathrm{E}_{Z_j \cap \mathrm{P}_{\mathrm{bdd}}\cap S^{c}}^{\mathrm{a.c}}(\mathbb{U}_{\omega})  \leq 2j.
\end{equation}
So, combining (\ref{eq_ishii_1}) and (\ref{eq_ishii_2}), we have finally proven that
$$\mathrm{rk}\ \mathrm{E}_{Z_j}^{\mathrm{a.c}}(\mathbb{U}_{\omega}) \leq 2j,$$
which achieve the proof of the theorem.
\qed

We recall that $\nu_1$ denote the Haar measure on $\SM^1$. We can deduce from Theorem \ref{thm_ishii_pastur} the following corollary.

\begin{coro}\label{cor_ac_lyap}
If for $\nu_1$-almost every $z\in \SM^1$, $\gamma_1(z)\geq \cdots \geq \gamma_L(z) >0$, then $\Sigma_{\mathrm{ac}}=\emptyset$. 
\end{coro}

\proof
If $z\in \SM^1$ is such that $\gamma_L(z)>0$, then no Lyapunov exponent vanishes at $z$, which means that $z\in Z_0$. By Theorem \ref{thm_ishii_pastur}, $ \mathrm{E}_{Z_0}^{\mathrm{a.c}}(\mathbb{U}_{\omega})=0$. Now, if $\Delta$ is a Borelian subset of $\SM^1$, since by hypothesis $\nu_1(Z_0)=1$,
$$ \mathrm{E}_{\Delta}^{\mathrm{a.c}}(\mathbb{U}_{\omega})=  \mathrm{E}_{\Delta \cap Z_0}^{\mathrm{a.c}}(\mathbb{U}_{\omega})=0,\ \mbox{for } \mathbb{P}\mbox{-a.e } \omega\in \Omega.$$
It implies $\Sigma_{\mathrm{ac}}=\emptyset$ by definition of  $\Sigma_{\mathrm{ac}}$.
\qed
\vskip 3mm

Finally, from Corollary \ref{cor_ac_lyap} and Theorem \ref{thm_pos_Lyap_intro} we deduce immediately Theorem \ref{thm_ishii_pastur_intro}.
\vskip 6mm

\noindent \textbf{Acknowledgements.} Authors are grateful for the support of the GDR DYNQUA which allow H. Boumaza to stay for a week at the Institut Joseph Fourier from Grenoble were part of this research was done. They would also like to thank Alain Joye for fruitful discussions.

\vskip 1cm

\end{document}